\documentclass[11pt]{article}

\usepackage{amssymb}
\usepackage{amsmath}
\usepackage{vmargin}
\usepackage{graphicx}
\usepackage[tight]{subfigure}
\usepackage{ifpdf}
\usepackage{hyperref} 

\setmargrb{.6in}{.3in}{.6in}{.4in}

\newcommand{\order}{\mathcal{O}}

\newcommand{\GeV}{\ \mathrm{GeV}}

\newcommand{\Mgrav}{M_*}
\newcommand{\MGUT}{M_\mathrm{GUT}}

\newcommand{\MR}{M_R}
\newcommand{\tb}{\tan\!\beta}
\newcommand{\cb}{\cos\!\beta}
\newcommand{\tmg}{\tau \rightarrow \mu \gamma}
\newcommand{\teg}{\tau \rightarrow e \gamma}
\newcommand{\meg}{\mu \rightarrow e \gamma}
\newcommand{\bsg}{B \rightarrow X_s \gamma}

\newcommand{\bsbsbar}{$B_s$--$\overline{B_s}$}

\newcommand{\SphiK}{S_{CP}^{\phi K}}

\newcommand{\deltad}{\delta^d}
\newcommand{\deltal}{\delta^l}
\newcommand{\ded}[2]{(\deltad_{#1})_{#2}}
\newcommand{\del}[2]{(\deltal_{#1})_{#2}}

\newcommand{\incgr}[2][]{\IfFileExists{../pr-tvmet/KNLO/HFAG/#2.pdf}{%
    \includegraphics[#1]{../pr-tvmet/KNLO/HFAG/#2.pdf}}{%
    \includegraphics[#1]{#2.pdf}}}

\newcommand{\hepph}[1]{hep-ph/#1}
\newcommand{\hepex}[1]{hep-ex/#1}
\newcommand{\arXivid}[1]{arXiv:#1}

\begin{document}

\title{$B_s$ mixing phase and lepton flavor violation in supersymmetric SU(5)}
\author{Jae-hyeon Park\thanks{jae-hyeon.park@pd.infn.it}\\
  INFN, Sezione di Padova, via F Marzolo 8, I--35131, Padova, Italy
  \and
  Masahiro Yamaguchi\thanks{yama@tuhep.phys.tohoku.ac.jp}\\
  Department of Physics, Tohoku University, Sendai 980--8578, Japan}
\date{28 September 2008}
\maketitle

\begin{flushright}
\raisebox{2.9in}[0mm][0mm]{TU-813}
\end{flushright}
\vspace{-3.5em}

\begin{abstract}
  We inspect consequences of the latest $B_s$ mixing phase measurements
  on lepton flavor violation in a supersymmetric SU(5) theory.
  The $\order(1)$ phase, preferring a non-vanishing squark mixing,
  generically implies $\tau \rightarrow (e + \mu)\,\gamma$ and $\meg$.
  Depending on the gaugino and the scalar mass parameters as well as $\tb$,
  the rates turn out to be detectable
  or even already excessive,
  if the $RR$ mass insertion of down-type squarks is nonzero.
  We find that it becomes easy
  to reconcile $B_s$ mixing phase with lepton flavor violation
  given:
  gaugino to scalar squared mass ratio around $1/12$,
  both $LL$ and $RR$ insertions with decent sizes, and
  low $\tb$.
\end{abstract}



In the last few years, two experiments at Tevatron have been accumulating
information on the mixing of the $B_s$-meson.
The precision of the mass splitting $\Delta M_s$
between the two mass eigenstates
composed of $B_s$ and $\overline{B_s}$, by now has reached
the level of 0.7\% \cite{Abulencia:2006ze},
which is comparable to that of the $B_d$-meson \cite{Barberio:2008fa}.
Despite its high accuracy,
$\Delta M_s$ is not showing any incompatibility with the Standard Model (SM)\@.
This should be regarded as yet another triumph of the model.
However, a point to keep in mind at the present moment is that
it is not easy to separate an extra contribution within $\Delta M_s$,
even if one exists, from that of the SM,
due to the large theoretical uncertainty around 30\%
stemming from errors
in the $\Delta B = 2$ hadronic matrix element and
the Cabibbo--Kobayashi--Maskawa (CKM) quark mixing matrix \cite{Lenz:2006hd}.

On the other hand, the mixing phase, denoted by $\phi_s$, does not suffer from
these theoretical uncertainties, and one can make a closer
connection between its data and a theory possibly involving new physics
\cite{Fleischer:2008uj}.
Let us choose the notation $\phi_s$ to represent what is called
$\phi_s^{J/\psi\phi}$ by the Heavy Flavor Averaging Group (HFAG)\@
\cite{Barberio:2008fa}.
In the SM, one has $\phi_s \simeq - 2 \eta \lambda^2 \simeq -0.04$.
On the experimental side, it is still much less precise
than $\sin 2\beta$.
Nonetheless, $\phi_s$ is already becoming a useful probe
into the flavor sector of an extension of the SM\@.
In particular, one could observe an interesting tendency
in both data from D\O\ \cite{Abazov:2008fj} and CDF \cite{Aaltonen:2007he},
that each result appeared to favor a negative $\order(1)$ value of $\phi_s$.
This tendency came to stand out after the UTfit collaboration,
based on the two experiments, reported that their global fit
showed a $3.7\ \sigma$ discrepancy of $\phi_s$ from its SM value
\cite{Bona:2008jn}.
This deviation, however, has decreased to $2.5\ \sigma$
after they updated their analysis including newly available
experimental information from D\O\ \cite{UTfit update}.
The latest constrained fit by HFAG shows that \cite{Barberio:2008fa}
\begin{equation}
  \label{eq:phis hfag}
\phi_s = -0.76^{+0.37}_{-0.33} \text{ \ or \ } \mbox{$-2.37$}^{+0.33}_{-0.37} ,
\end{equation}
which is consistent with the SM at the level of $2.4\ \sigma$.
Still, it is too early to draw a definite conclusion.
If the difference solidifies, it should be a clean
indication of a new source of $CP$ violation.

A supersymmetric extension of the SM has potential new sources of
flavor and/or $CP$ violation in its soft supersymmetry breaking terms.
It might be conceivable that one of them is revealing its existence
through the above anomaly.
We employ the notation of mass insertion parameters,
written in the form of $\ded{ij}{AB}$ with
the generation indices $i, j = 1,2,3$ and the chiralities $A,B = L,R$.
We do not only use their usual definition at the weak scale
\cite{Hall:1985dx},
but also borrow the same notation to specify an off-diagonal element
of the soft scalar mass matrix at $\MGUT$,
the unification scale \cite{fcncgut}.
For instance, we define
$\ded{ij}{LL} \equiv [m^2_q]_{ij} / \widetilde{m}^2$ at $\MGUT$,
where $\widetilde{m}^2$ is the averaged diagonal entry of $m^2_q$,
the soft scalar mass matrix of the SU(2) doublet squarks
in the basis where the down-type quark Yukawa matrix is diagonal.
Being a transition between the second and the third families,
\bsbsbar\ mixing is naturally associated with $\ded{23}{AB}$.
Among the four possibilities,
the $LR$ and the $RL$ mass insertions tend to cause
an unacceptable change in $\bsg$
before they can give an appreciable modification
to \bsbsbar\ mixing \cite{LRRLbsgam}.
Therefore, we focus on $LL$ and $RR$ mixings in what follows.

One could think of a more interesting situation
by working with a grand unified theory (GUT)\@.
Since a single GUT multiplet contains both (s)quarks and (s)leptons,
flavor transitions in the two sectors are related 
\cite{Baek:2001kh,Hisano:2003bd,Ciuchini:2003rg,%
Dutta:2006gq,Cheung:2007pj,Ciuchini:2007ha,Borzumati:2007bn,%
Goto:2007ee,Hisano:2008df}.
Then, one immediately arrives at the
conclusion that the new source of $b \leftrightarrow s$ transition,
needed to account for $\phi_s$,
generically implies lepton flavor violation (LFV) \cite{Parry:2007fe}.
We wish to consider this scenario in a model independent fashion
taking SU(5) as the unified gauge group.
This work has at least two differences from the article just referenced.
First, we take into account the running effects of squark masses
below $\MGUT$.
The diagonal components of the squark and the slepton mass matrices
grow in the course of running, and this effect is more important
to squarks than to sleptons due to the gluino mass contribution.
Because of this difference, the gaugino to scalar mass ratio
at $\MGUT$ plays an important role in determining
relative strengths of the two types of flavor violations,
hadronic and leptonic.
This finding will be demonstrated later in the results.
Second, we inspect additional observables
such as $\meg$, $\SphiK$, and neutron electric dipole moment (EDM)\@.
In addition to $\tmg$, $\meg$ turns out to be highly sensitive to
$\ded{23}{RR}$ thanks to $\del{13}{RR}$ that is radiatively generated from
top Yukawa coupling and CKM mixing
\cite{fcncgut}.


In a related work \cite{fcncgut},
we present a more detailed study on supersymmetric flavor violation
in a SU(5) GUT\@.
Let us recapitulate highlights thereof,
relevant to the following discussions.
The first topic is the connection of a leptonic process to a squark mixing.
We ignore the running effects on slepton mixings
from neutrino Yukawa couplings below $\MGUT$.
In cases where there are sizeable
right-handed down-type squark mixings, 
they lead to LFV decays dominated by chargino loops.
If one has a perfect alignment between the mass eigenstates of
quarks and leptons,
$\ded{ij}{RR}$ implies the transition of $l_j \rightarrow l_i$.
However, this straightforward correspondence may be broken by
the inclusion of non-renormalizable terms into the superpotential
as a solution to the wrong quark--lepton mass relations of the lighter two
families.
With the assumption that the cutoff scale of the GUT is
two orders of magnitude higher than $\MGUT$,
one can nevertheless have
\begin{equation}
  \label{eq:brtemg}
  B (\tau \rightarrow (e + \mu)\,\gamma) \propto
  |\del{13}{LL}|^2 + |\del{23}{LL}|^2 \approx
  |{\ded{13}{RR}}|^2 + |{\ded{23}{RR}}|^2 +
  \order[\cos^2\!\beta\, (\deltad_{RR})^2] ,
\end{equation}
in terms of insertions at $\MGUT$,
exploiting the fact that the breakdown of $b$--$\tau$ alignment
is suppressed by $\cb$ \cite{Baek:2001kh}.
Therefore, non-vanishing $\ded{23}{RR}$ causes either $\tmg$ or $\teg$
\cite{fcncgut}.
A tau decay may be linked also to the left-handed squark mixings.
One can reuse~(\ref{eq:brtemg}) except that each chirality index should be
flipped to the opposite one.
Another difference is that the process amplitude is dominated by
a neutralino loop, and thus is much smaller than one from a chargino loop,
for a given size of mixing.

An analogous statement can be made regarding $\meg$, albeit
in a somewhat involved form.
It is applicable only to the $RR$ mixings,
due to the mechanism by which the decay occurs.
The branching fraction has a lower bound such that
\begin{equation}
  B (\meg)  \gtrsim
  k \times
  \min \{ |\del{13}{RR}|^2, |\del{23}{RR}|^2 \} \cdot
  [ |{\ded{13}{RR}}|^2 + |{\ded{23}{RR}}|^2 ] ,
\end{equation}
with the terms suppressed by $\cos^2\!\beta$ omitted.
The proportionality constant $k$ can be worked out by
calculating the rate from a neutralino loop with triple mass insertions
$\del{13}{RR}\del{33}{RL}\del{32}{LL}$.
The second factor is at least around $\ded{13}{LL}$
which is supposed to have received radiative corrections
from top Yukawa coupling and CKM mixing \cite{Barbieri:1994pv}.
Thus, nonzero $\ded{23}{RR}$ gives rise to $\meg$,
unless there is a fine-tuning among parameters in the superpotential
and the soft supersymmetry breaking sector \cite{fcncgut}.

The second topic, on hadronic processes,
is the competition between squark decoupling and the growth of
a $\Delta$ parameter,
as the diagonal components of the squark mass matrix increase.
By $\Delta$, we mean the off-diagonal part of a sfermion mass matrix.
Suppose that the $\delta$ parameters defined above and
the gaugino mass $M_{1/2}$ are fixed at $\MGUT$.
Imagine that one can increase
$m_0$, the common diagonal entries of soft squark mass matrix at $\MGUT$,
from the value which make the gluino and the squark masses coincide
at the weak scale.
This value corresponds to $x \equiv M_{1/2}^2 / m_0^2 \approx 0.7$.
As $m_0$ increases, $(\Delta^d_{ij})_{AB} \equiv m_0^2 \times \ded{ij}{AB}$
grows as well, thereby exerting more and more influence
on low energy flavor violation such as $B_s$ mixing.
At some point, however, squark loop effects begin to decouple
as the squarks become too heavy.
For \bsbsbar\ mixing, this is around $x = 1/12$.
This gaugino to scalar mass ratio could be regarded as a condition
for optimizing the sensitivity of a hadronic process to
flavor non-universality at $\MGUT$ \cite{fcncgut}.
The importance of this observation is more pronounced when one tries
to compare hadronic and leptonic constraints since the latter
is monotonically weakened as $m_0$ is being raised.

Having briefed the reader on qualitative aspects of flavor physics
in a supersymmetric GUT,
we proceed to computation.
We take the same procedure of numerical analysis as in Ref.~\cite{fcncgut}.
As was already mentioned,
we restrict ourselves to $LL$ and $RR$ mixings of down-type squarks.
Regarding patterns of the two insertions, we consider three scenarios:
the $LL$ scenario, the $RR$ scenario, and the $LL=RR$ scenario.
The meaning of each name should be self-explanatory except that
we set an $LL$ insertion, unless it is a scanning variable,
to a value generated by renormalization group (RG) running
from the supersymmetry breaking mediation scale $\Mgrav$
down to $\MGUT$, where
$\Mgrav$ is taken to be the reduced Planck scale.
We do this for $\ded{12}{LL}$ and $\ded{13}{LL}$ as well as $\ded{23}{LL}$.
These boundary conditions are given at $\MGUT$
with which we solve one-loop RG equations down to the weak scale.
We consider only the gluino loop contributions to a quark sector process.
We display the portion of the parameter space
permitted by each constraint
on the complex plane of a GUT scale mass insertion.
As for $\phi_s$, we use the 90\% confidence level (CL) region from
HFAG \cite{Barberio:2008fa},
\begin{equation}
  \phi_s \in [-1.26, -0.13] \cup [-3.00, -1.88] .
\end{equation}
For concreteness, we assume that there is an
exact quark--lepton flavor alignment.
Regarding $\tmg$,
it is straightforward to translate their bounds presented below
to a case with quark--lepton misalignment discussed above---%
interpret $B (\tmg)$ as $B (\tau \rightarrow (e + \mu)\,\gamma)$.
This prescription is applicable to all the three scenarios considered
in this work.
As for $\meg$, barring accidental cancellations,
a contour does not need a modification
in the $RR$ and $LL=RR$ scenarios, while
we do not have a systematic way to account for a misalignment
in the $LL$ scenario.
We will elaborate on this later on. 
In order to demonstrate the role of the gaugino to scalar mass ratio,
we fix $M_{1/2} = 180\GeV$, which makes the gluino mass be $500\GeV$
at the weak scale, and then try two different values of
$m_0 = 220\GeV$ and $600\GeV$,
corresponding to the right-handed down-type squark masses of
$500\GeV$ and $750\GeV$ at the weak scale, respectively.
The former $m_0$ results in
a benchmark case often considered in the literature,
and the latter $m_0$
optimizes the sensitivity of neutral meson mixing
to $\delta$'s at the GUT scale.
We also vary $\tb$ from 5 to 10.
Other details can be found in Ref.~\cite{fcncgut}, such as
experimental inputs in use and ways to impose them
as constraints.

\begin{figure}
  \centering
  \subfigure[$m_0 = 220 \GeV,\ M_{1/2} = 180 \GeV,\ \tb = 5$]{\incgr[height=63mm]{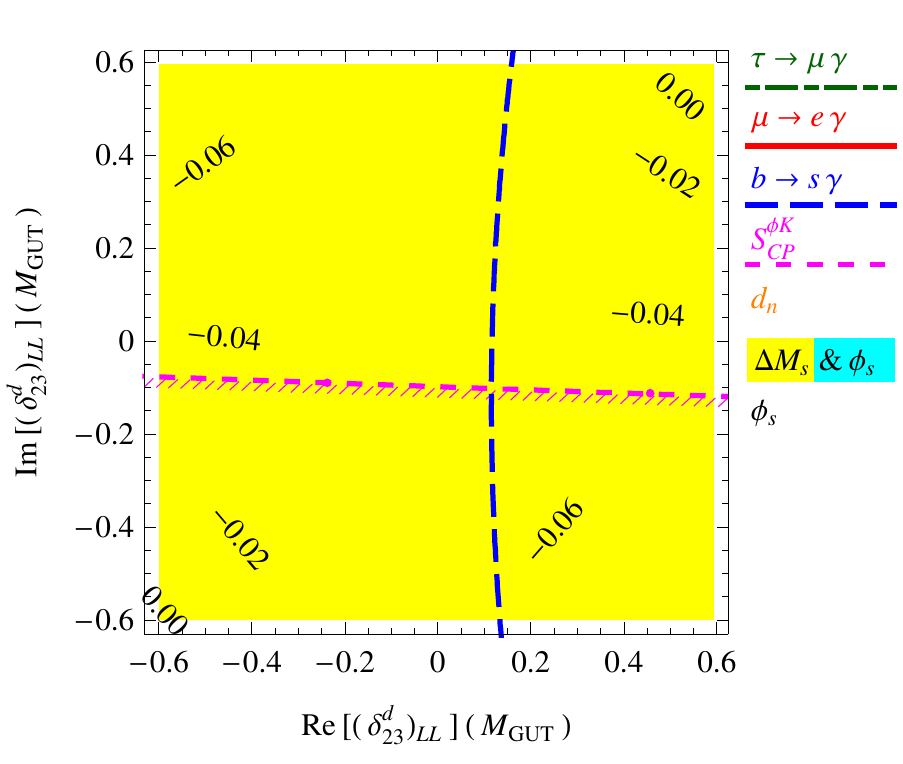}}
  \subfigure[$m_0 = 220 \GeV,\ M_{1/2} = 180 \GeV,\ \tb = 10$]{\incgr[height=63mm]{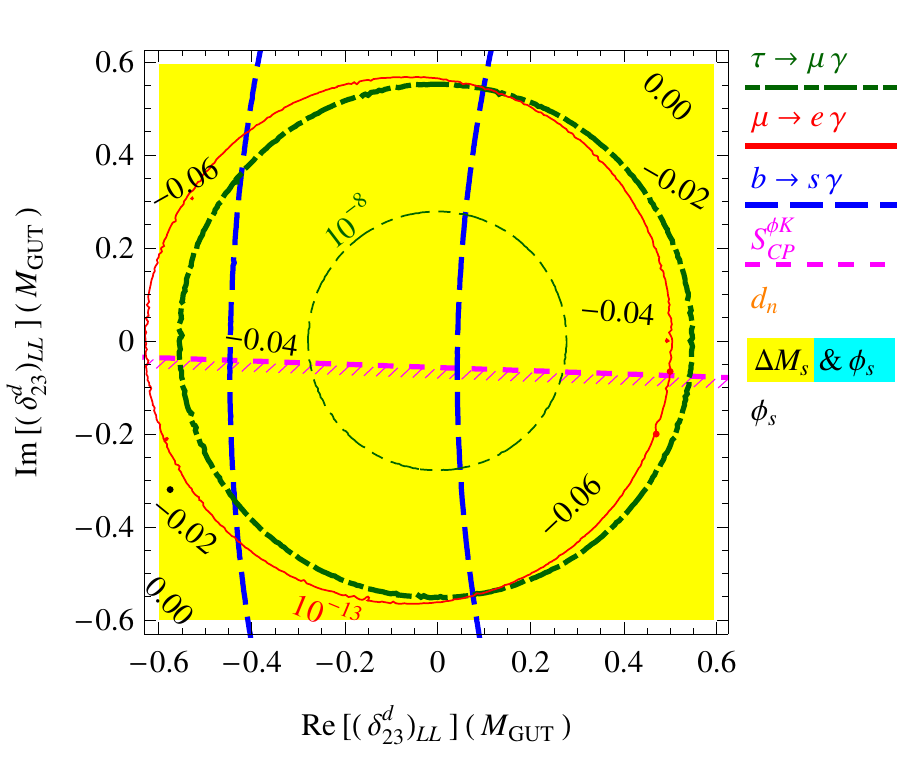}}
\\
  \subfigure[$m_0 = 600 \GeV,\ M_{1/2} = 180 \GeV,\ \tb = 5$]{\incgr[height=63mm]{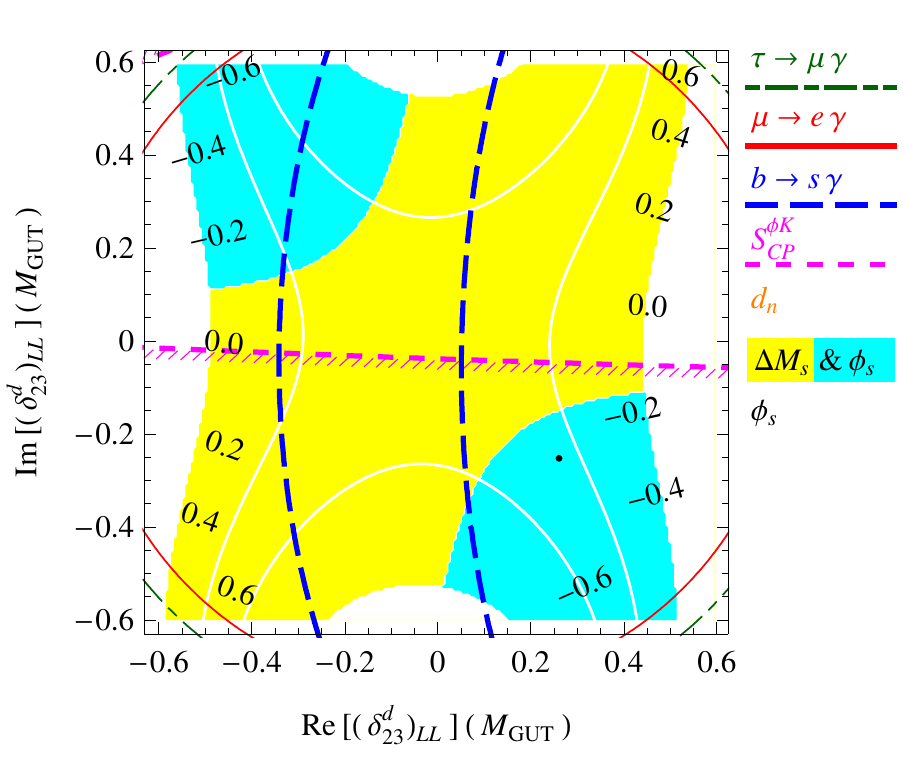}}
  \subfigure[$m_0 = 600 \GeV,\ M_{1/2} = 180 \GeV,\ \tb = 10$]{\incgr[height=63mm]{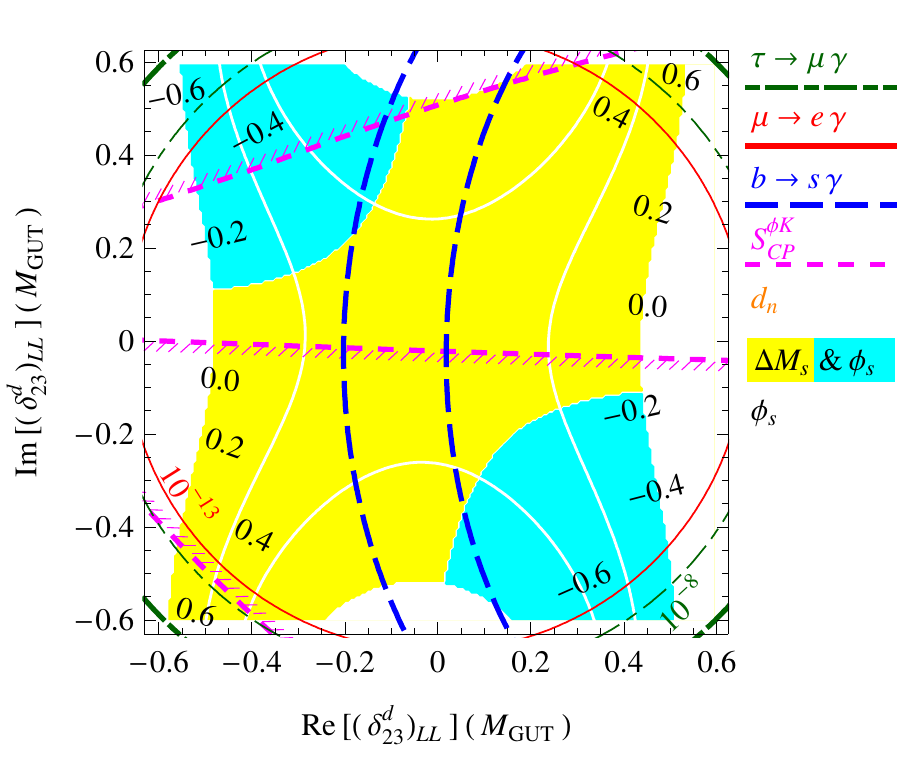}}
  \caption{Constraints on the complex plane of $\ded{23}{LL}$,
    with $\ded{12}{LL}$ and $\ded{13}{LL}$ generated from
    RG running between the reduced Planck scale and the GUT scale.
    For $\tmg$, the thick circle is the current upper bound,
    and the thin circle is an upper bound
    from the prospective branching ratio limit, $10^{-8}$.
    For $\meg$, the thin circle shows the projected bound
    on the branching ratio, $10^{-13}$.
    A light gray (yellow) region is allowed by $\Delta M_s$,
    given 30\% uncertainty in the $\Delta B = 2$ matrix element,
    and a gray (cyan) region is further consistent with $\phi_s$.
    The white curves 
    mark a possible improved
    constraint from $\Delta M_s$
    with 8\% hadronic uncertainty.
    Of the two sides of the $\SphiK$ curve,
    the excluded one is indicated by thin short lines.}
  \label{fig:23LL}
\end{figure}

First, let us examine the $LL$ mixing scenario.
The region preferred by each process is shown in Figs.~\ref{fig:23LL}.
Among the four figures, Figs.~(a) and (b) are for lower $m_0$.
For this $m_0$, one recognizes that
the supersymmetric contribution to \bsbsbar\ mixing
is not enough to fit the $\phi_s$ data
even if one allows for an $\order(1)$ mass insertion.
The dotted contour lines tell us that
a maximal alteration in $\phi_s$ that can be expected is about $0.1$.
They reveal that the other experimental constraints are not
the primary reason why the $LL$ mixing scenario with lower $m_0$
is inadequate for making an $\order(1)$ change in $\phi_s$.
The mixing is simply unable to make an enough difference,
due to the dilution of squark mixing
by gluino mass contribution in the course of RG running
down to the weak scale.
In Figs.~(c) and (d), one can find gray (cyan) regions that
lead to $\phi_s$ within its 90\% CL intervals.
They involve an $\order(1)$ mass insertion
between the second and the third families
of left-handed squarks.
For this value of $m_0$, squark mixing given at $\MGUT$ is less diluted
by the running of diagonal components of the mass matrix.
However, the supersymmetric disturbance is not only enhanced in $B_s$ mixing,
but also in $\bsg$.
Because of this, the bulk of a gray zone is excluded by $\bsg$.
The disturbance in this decay mode grows with $\tb$
\cite{Gabbiani:1988rb,Ko:2008xb}.
One can see that the $\bsg$ constraint is severer in Fig.~(d) than in Fig.~(c),
and that there remains a bigger viable corner for lower $\tb$.
Also note that for $\tb = 10$,
the phase of $\ded{23}{LL}$, needed to fit $\phi_s$,
modifies $\SphiK$ so that it goes out of its current $2\sigma$ range,
except in a small part of the gray area.
Some parts of the regions favored by $\phi_s$,
give rise to $\tmg$ and/or $\meg$ so much as they can be detected
at a super $B$ factory \cite{Hashimoto:2004sm,Bona:2007qt} or
the MEG experiment \cite{MEG}.
LFV rates in those parts increase with $\tb$
enlarging their discovery chance,
although large $\tb$ is disfavored by $\bsg$ and $\SphiK$.
Remember that the displayed LFV branching ratios have been calculated
under the assumption that the quark and the lepton mass eigenstates are
aligned.
We will come back to consequences of relaxing this assumption later.

\begin{figure}
  \centering
\subfigure[$m_0 = 220 \GeV,\ M_{1/2} = 180 \GeV,\ \tb = 5$]{\incgr[height=63mm]{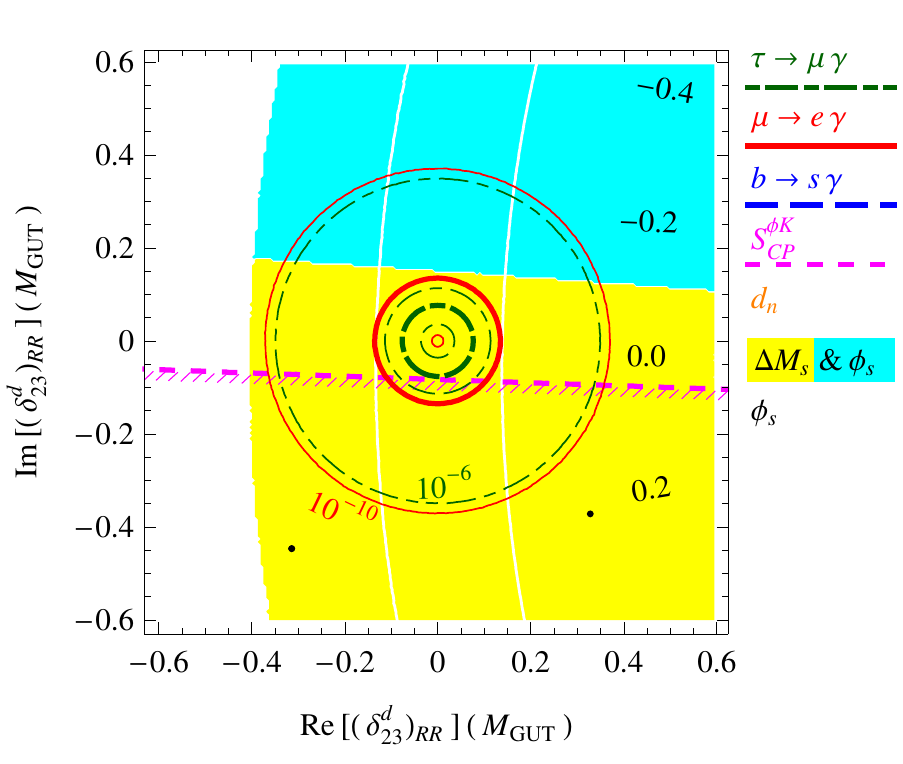}}
\subfigure[$m_0 = 220 \GeV,\ M_{1/2} = 180 \GeV,\ \tb = 10$]{\incgr[height=63mm]{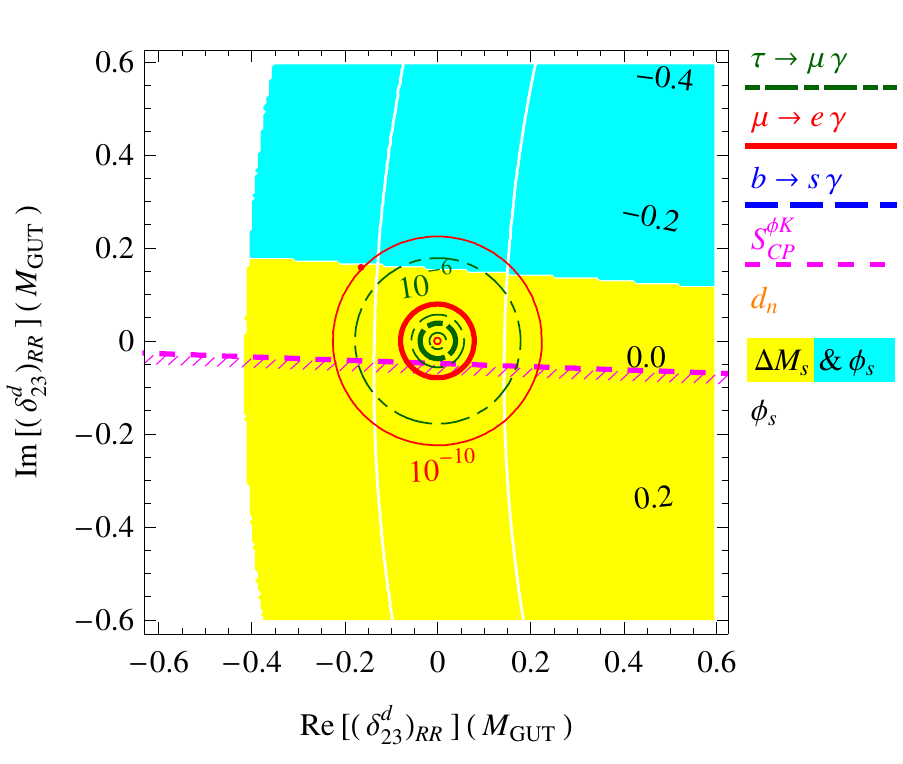}}
\\
\subfigure[$m_0 = 600 \GeV,\ M_{1/2} = 180 \GeV,\ \tb = 5$]{\incgr[height=63mm]{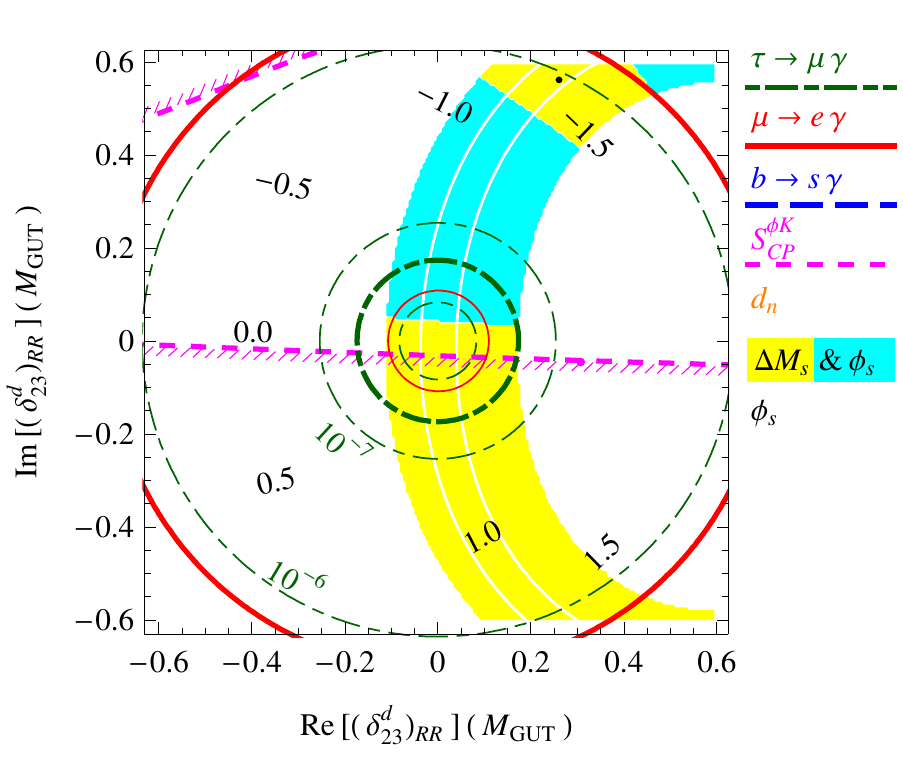}}
\subfigure[$m_0 = 600 \GeV,\ M_{1/2} = 180 \GeV,\ \tb = 10$]{\incgr[height=63mm]{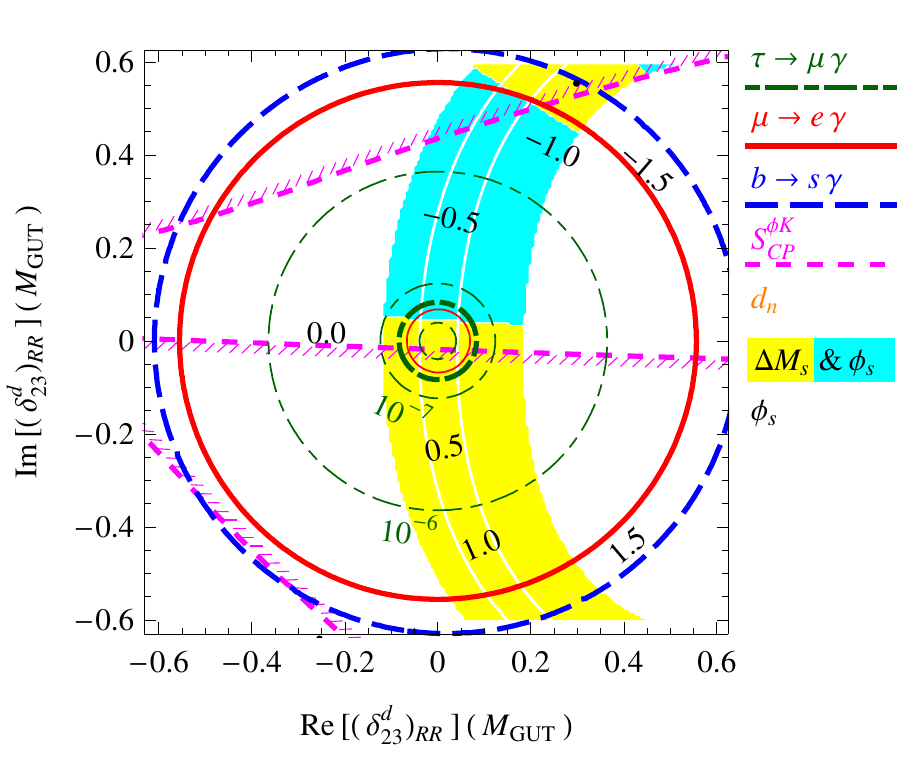}}
  \caption{Constraints on the complex plane of $\ded{23}{RR}$,
    with $\ded{ij}{LL}$ generated from
    RG running between the reduced Planck scale and the GUT scale.
    For $\tmg$, the thick circle is the current upper bound,
    and the thin circles are, from inside, branching ratios of
    $10^{-8}$, $10^{-7}$, $10^{-6}$, respectively.
    For $\meg$, the thick circle is the current upper bound,
    and the thin circles are, from inside, branching ratios of
    $10^{-13}$, $10^{-10}$, respectively.
    A light gray (yellow) region is allowed by $\Delta M_s$,
    given 30\% uncertainty in the $\Delta B = 2$ matrix element,
    and a gray (cyan) region is further consistent with $\phi_s$.
    The white curves running from top to bottom mark a possible improved
    constraint from $\Delta M_s$
    with 8\% hadronic uncertainty.
    Of the two sides of the $\SphiK$ curve,
    the excluded one is indicated by thin short lines.}
  \label{fig:23RR}
\end{figure}

Let us turn to the $RR$ scenario.
The plots are presented in Figs.~\ref{fig:23RR}.
Comparing the first two of Figs.~\ref{fig:23RR} with
those of Figs.~\ref{fig:23LL}, one can notice that
gray (cyan) regions are visible here, unlike the $LL$ scenario.
This is due to the $LL$ insertion induced by RG running
from $\Mgrav$ down to the weak scale \cite{AlvarezGaume:1981wy}.
The presence of $\ded{23}{LL}$ enhances the effect
of $\ded{23}{RR}$ on \bsbsbar\ mixing
\cite{Gabbiani:1988rb,Ko:2008xb,23LL23RR}.
However, those regions leading to $\phi_s$
within its 90\% CL range,
are excluded by the current bounds from $\tmg$ \cite{Hayasaka:2007vc}
and $\meg$ \cite{Brooks:1999pu},
even for $\tb$ as low as 5.
It seems to be hard to satisfy both $\phi_s$ and LFV
with an $RR$ insertion with low $m_0$.
This should be contrasted with the $LL$ scenario
where LFV was not a major problem.
Given non-vanishing $\ded{23}{RR} (\MGUT) = \del{23}{LL}^* (\MGUT)$,
$\tmg$ is dominated by a chargino loop,
while for mass insertions with the opposite chiralities,
it occurs through a neutralino loop.
A chargino loop contributes a larger
amplitude per mass insertion than a neutralino loop \cite{Hisano:1995cp}.
Therefore, $\tmg$ acts as a tighter restriction here than in the $LL$ scenario.
Note that $\meg$ occurs as well.
This stems from the nonzero $\del{13}{RR}$ set as a boundary condition
at $\MGUT$.
This value is expected from
the radiative correction from top Yukawa coupling and CKM mixing.
Picking up this insertion in addition to $\del{33}{RL}$ and $\del{32}{LL}$,
a neutralino loop for $\meg$ can be completed,
which is enhanced by the factor $m_\tau / m_\mu$ coming from $\del{33}{RL}$
\cite{Baek:2001kh,Hisano:1995cp,megtriplefromGUT,Paradisi:2005fk}.
Neutron EDM, denoted by $d_n$,
restricts the imaginary part of $\ded{23}{RR}$
with the aid of $\ded{23}{LL}$ \cite{neutronEDM}.
As a result, $d_n$ is setting a limit to which an $RR$ insertion
can satisfy $\phi_s$, although it is weaker than LFV\@.
Contrasting Figs.~\ref{fig:23RR}~(a) and (b) shows that
both LFV and $d_n$ constraints become tighter for higher $\tb$.
Next, we switch to a higher value of $m_0$.
Compared to the upper row with lower $m_0$,
the cases in Figs.~(c) and (d) need
a smaller size of mass insertion to give an enough contribution
to \bsbsbar\ mixing, to its phase in particular.
The reason has been already explained.
In contrast, LFV is suppressed because of heavier sleptons.
These two changes make
it easier to fit $\phi_s$ with smaller LFV rates.
Enhancement of hadronic processes, though, leads to a stricter $d_n$ limit.
A region allowed by $d_n$ and $\Delta M_s$ around the origin,
is separated from the $\phi_s$ region.
Recall that $d_n$ is influenced through the combination of
$\mathrm{Im} [\ded{23}{LL} \ded{23}{RR}^*]$.
Thus the band obeying $d_n$ can be rotated to overlap the gray region
by altering $\ded{23}{LL}$ at $\MGUT$.
This can happen if $\ded{23}{LL}$ is initially non-vanishing
with a complex phase at $\Mgrav$.
Alternatively, the non-renormalizable terms in the superpotential could 
alter the insertion while it runs from $\Mgrav$ to $\MGUT$.
The presented plots are valid for the phase of $\ded{23}{LL}$ equal to
$\arg (- V_{ts}^* V_{tb})$.
It is noticeable that $\bsg$ is not playing a very important role.
Its branching ratio is not affected as much as in the $LL$ mixing scenario
since the supersymmetric amplitude does not interfere with the SM one.
Still, the $\bsg$ ring should be able to touch the gray region
for $\tb$ higher than 10.
LFV and $d_n$ are also enhanced for high $\tb$.
Therefore, lowering $\tb$ helps satisfy LFV and $d_n$ as well as $\phi_s$.
The contour lines of $\phi_s$ and a LFV branching fraction show
the correlation between them.
Suppose that $\tb = 5$.
One can find that the region preferred by $\phi_s$ involves
the $\tmg$ rate in the vicinity of the current upper limit.
For example, fitting the central value of $\phi_s$
in (\ref{eq:phis hfag}) causes
$B(\tmg)$ to be around $10^{-7}$ which is already ruled out by the Belle data
\cite{Hayasaka:2007vc}.
The area still surviving
could be explored by current and future experiments.
The magnitude of mass insertion accessible with the sensitivity of $10^{-8}$,
attainable at a super $B$ factory, is depicted by a thin circle
inside the current upper bound.
The prospect may be brighter according to Ref.~\cite{Bona:2007qt},
which proposes an upper limit of
$2\times 10^{-9}$ with $75\ \mathrm{ab}^{-1}$.
The gray region is also expected to bring about
$\meg$ at a rate that can be probed by MEG\@.
A more detailed model independent study on the connection between
$\ded{23}{RR}$ and $\meg$ has been performed in Ref.~\cite{fcncgut}.

The preceding results are based on the supposition
that the quark and the lepton mass eigenstates are aligned to each other.
With the following modifications,
they can be applied to cases where this alignment is disturbed
by the Planck-suppressed non-renormalizable operators incorporated
to reproduce masses of the lighter quarks and leptons.
Replace $\tmg$ by $\tau \rightarrow (e + \mu)\,\gamma$, i.e.\
interpret the branching ratio of the former as that of the latter.
Obtain a new thick $\tau \rightarrow (e + \mu)\,\gamma$ ring by
expanding the old thick $\tmg$ ring, in order to
encompass the events of $\teg$ \cite{Aubert:2005wa}.
For this, multiply the old radius by 1.9.
Leave the thin $\tau \rightarrow (e + \mu)\,\gamma$
(namely former $\tmg$) circles untouched.
The $\meg$ contours should be kept as they are
in the $RR$ scenario, and discarded in the $LL$ scenario.
The net consequence of these operations is that
the current upper bound from $\tmg$ has been relaxed by the factor of 1.9
and $\meg$ has been disconnected from the $LL$ mixing.
We come back to the plots of $RR$ scenario in Figs.~\ref{fig:23RR}.
For lower $m_0$,
the conflict between LFV and $\phi_s$ is not very much ameliorated,
partly due to the still-strong $\tau \rightarrow (e + \mu)\,\gamma$
and partly due to $\meg$.
For higher $m_0$, the overlap broadens between
the zones preferred by LFV and $\phi_s$.
One can read off
the correlation between $\phi_s$ and $B(\tau \rightarrow (e + \mu)\,\gamma)$
from their contour lines.
Given a prediction of $B(\tau \rightarrow (e + \mu)\,\gamma)$ and
the same future branching ratio reaches of $\tmg$ and $\teg$, say $10^{-8}$,
the chance of discovering either is minimized when
the two modes have equal rates.
Even in this worst case, a point on the plot could be probed by LFV
if it leads to $B(\tau \rightarrow (e + \mu)\,\gamma)$
above $2\times 10^{-8}$.
Indeed, one can find a substantial part of a gray region
with this property.
Next, we reinterpret Figs.~\ref{fig:23LL} of the $LL$ scenario.
There, the role of LFV was not outstanding already before modification.
Now, imposing $\tau \rightarrow (e + \mu)\,\gamma$ instead of $\tmg$
moves its thick circle outside the visible range.
This makes the current LFV data further irrelevant to
an $\order(1)$ mixing.

Let us digress a little to remark on large neutrino Yukawa couplings.
If right-handed neutrinos are heavy,
the weak scale mass insertions of sleptons
receive corrections from the neutrino Yukawa couplings
while running below $\MGUT$ \cite{Borzumati:1986qx}. 
This contribution makes a shift in the position of
a LFV circle on the $RR$ mixing plot.
This might improve or worsen the
compatibility between LFV and $\phi_s$, depending on the direction
of the displacement.
There are cases with specific conditions where
one can easily guess the consequences.
Suppose that the scalar masses are universal at $\Mgrav$
and that the right-handed neutrinos are integrated out
at a single scale $M_R$.
In this case, $\ded{23}{RR}$, displayed in Figs.~\ref{fig:23RR},
is assumed to arise solely from neutrino Yukawa couplings.
Then a LFV upper bound shrinks by the factor,
$\ln (\Mgrav / \MGUT) / \ln (\Mgrav / \MR)$ \cite{fcncgut},
leaving a less room for $RR$ mixing at $\MGUT$
than is shown in Figs.~\ref{fig:23RR}.
One can also apply this method to a case where
there is a large hierarchy among the right-handed
neutrino masses, by replacing $\MR$
with the largest eigenvalue of $M_N$ \cite{Hisano:2003bd}.

\begin{figure}
  \centering
  \subfigure[$m_0 = 220 \GeV,\ M_{1/2} = 180 \GeV,\ \tb = 5$]{\incgr[height=63mm]{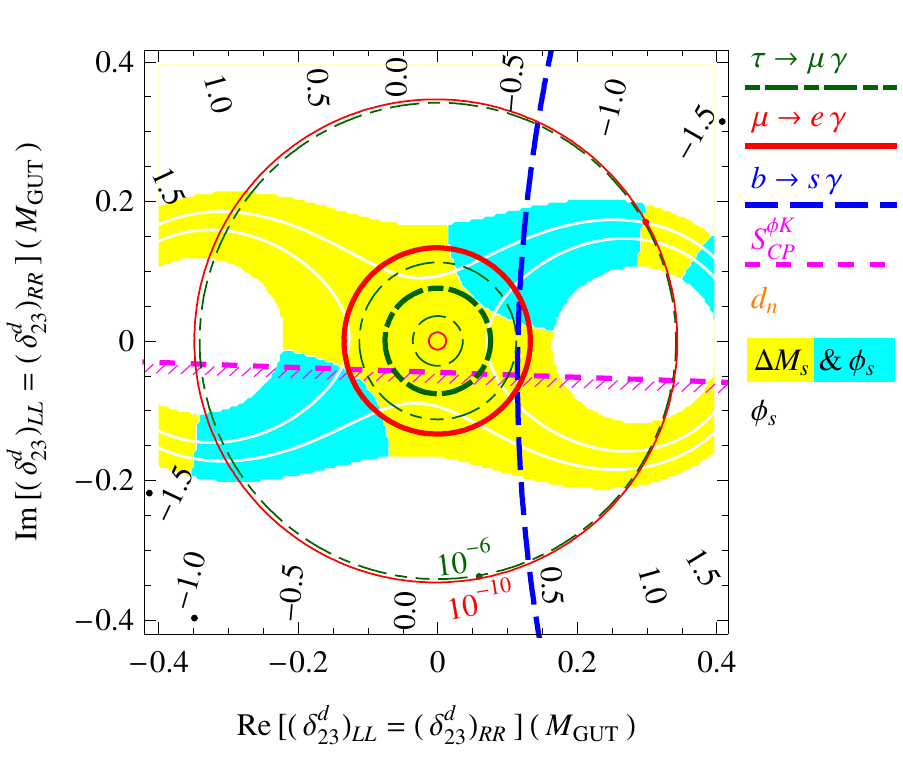}}
  \subfigure[$m_0 = 220 \GeV,\ M_{1/2} = 180 \GeV,\ \tb = 10$]{\incgr[height=63mm]{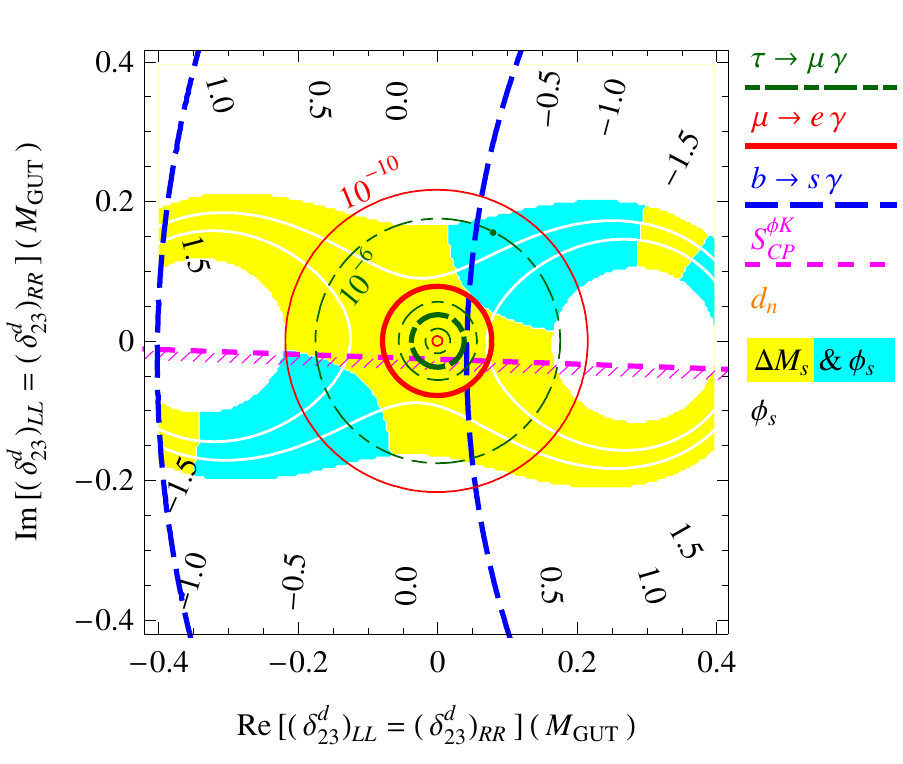}}
\\
  \subfigure[$m_0 = 600 \GeV,\ M_{1/2} = 180 \GeV,\ \tb = 5$]{\incgr[height=63mm]{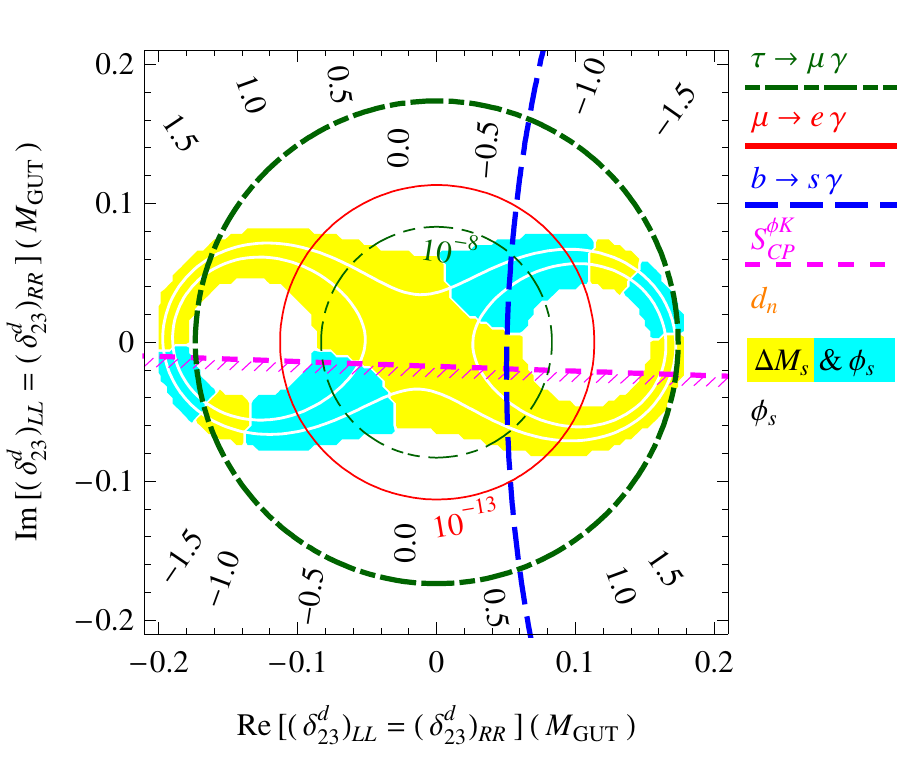}}
  \subfigure[$m_0 = 600 \GeV,\ M_{1/2} = 180 \GeV,\ \tb = 10$]{\incgr[height=63mm]{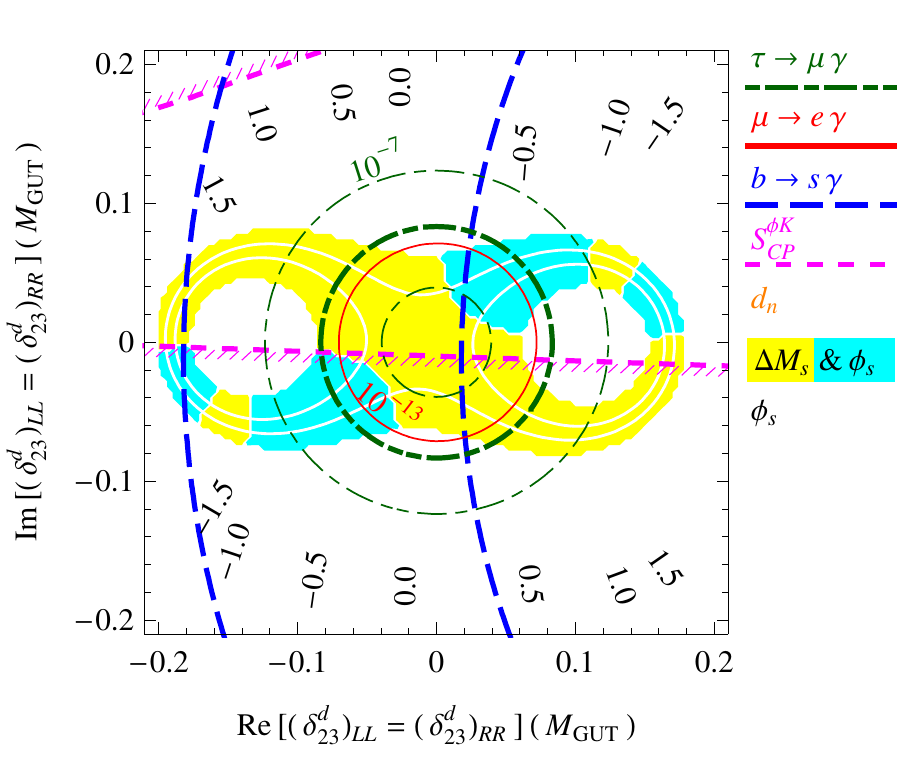}}
  \caption{Constraints on the complex plane of $\ded{23}{LL} = \ded{23}{RR}$,
    with $\ded{12}{LL}$ and $\ded{13}{LL}$ generated from
    RG running between the reduced Planck scale and the GUT scale.
    For $\tmg$, the thick circle is the current upper bound,
    and the thin circles are, from inside, branching ratios of
    $10^{-8}$, $10^{-7}$, $10^{-6}$, respectively.
    For $\meg$, the thick circle is the current upper bound,
    and the thin circles are, from inside, branching ratios of
    $10^{-13}$, $10^{-10}$, respectively.
    A light gray (yellow) region is allowed by $\Delta M_s$,
    given 30\% uncertainty in the $\Delta B = 2$ matrix element,
    and a gray (cyan) region is further consistent with $\phi_s$.
    The white curves 
    mark a possible improved
    constraint from $\Delta M_s$
    with 8\% hadronic uncertainty.
    Of the two sides of the $\SphiK$ curve,
    the excluded one is indicated by thin short lines.}
  \label{fig:23LL=23RR}
\end{figure}

We examine the last scenario with the condition that
$\ded{23}{LL} = \ded{23}{RR}$ at $\MGUT$.
The results are shown in Figs.~\ref{fig:23LL=23RR}.
Comparing Figs.~\ref{fig:23LL=23RR}~(a) and (b)
with Figs.~\ref{fig:23RR}~(a) and (b),
it appears that the conflict between LFV and $\phi_s$
has been much reduced here.
Simultaneous presence of $LL$ and $RR$ mixings
supplies a reinforced contribution to the $B_s$ mixing
even with a smaller size of each insertion about 0.2.
On the other hand, the LFV bounds remain almost the same
since the dominant source of each mode is $\ded{23}{RR}$,
as should be evident from Figs.~\ref{fig:23LL} and \ref{fig:23RR}.
Nonetheless, the LFV data shows a disagreement with $\phi_s$
for lower $m_0$, which grows severer for higher $\tb$.
Again, raising $m_0$ to the optimal point, one can enhance
supersymmetric effects on \bsbsbar\ mixing while suppressing LFV\@.
Especially, Fig.~\ref{fig:23LL=23RR}~(c) shows regions well inside
the LFV bounds which lead to $\phi_s$ in perfect agreement with
the latest global fit.
Part of those regions can satisfy $\SphiK$ and $d_n$ as well.
Notice that even though $d_n$ is very sensitive to
the product of $\ded{23}{LL}$ and $\ded{23}{RR}$,
it is not particularly enhanced relative to that in Figs.~\ref{fig:23RR},
where the $LL$ insertion is much smaller than here.
This is because $d_n$ is a function of
$\mathrm{Im} [\ded{23}{LL} \ded{23}{RR}^*]$ at the weak scale
and the contribution to this imaginary part arises only through
the RG-generated part of $\ded{23}{LL}$.
However, these two insertions both of large sizes can generically
disturb $d_n$ to a great extent,
once one relaxes the assumption that the phases
of the $LL$ and the $RR$ insertions are aligned.
This should be taken into account when
one tries to guess a situation with two uncorrelated large insertions.
On the other hand, it is always possible to escape from $d_n$
if one is willing to tune the relative phase between $\ded{23}{LL}$
and $\ded{23}{RR}$.
One finds that $B(\bsg)$ appears to prefer the left part of the plane.
This is because the SM value of the branching ratio, $3.2 \times 10^{-4}$,
that we use is smaller than the current central value from data
\cite{Barberio:2008fa}.
However, there is an enough possibility for the band to be shifted
left or right according to the other contributions from loops
involving chargino or charged Higgs.
Taking only the (half) width of the band as a criterion
for an acceptable size of mass insertion,
one could regard regions on both sides of the $\bsg$ curve as acceptable.
An area preferred by $\phi_s$ gives rise to $B(\tmg)$ around $10^{-8}$.
The rate of $\meg$ expected from the same area
is around the sensitivity of MEG\@.
In Fig.~(d), we vary $\tb$ up to 10.
The LFV bounds become tighter.
Nevertheless, there are corners of the gray zones that
obey all the constraints.
Remember that $d_n$ can be loosened by modifying the relative phase between
$\ded{23}{LL}$ and $\ded{23}{RR}$.
Obviously, the chance of observing LFV at a future experiment increases
with $\tb$.
Let us comment on more general cases with quark--lepton flavor misalignment.
One can convert each plot to a version for misalignment in the same way
as one did in the $RR$ scenario, since the LFV modes are dominated by
the $RR$ insertion in this scenario as well.
The maximum magnitude of insertion set by $\tmg$ should be multiplied by 1.9.
For lower $m_0$, most [Fig.~(a)] or all [Fig~(b)] of the region
favored by $\phi_s$ is still excluded by $\meg$
although it is a little weaker than $\tmg$ before the conversion.
For higher $m_0$, the conversion lifts the barrier of LFV even for $\tb = 10$,
thereby relieving the tension between $\phi_s$ and LFV\@.



Finally, we come to the summary.
We have assessed consequences of the latest $\phi_s$ data
on scalar flavor non-universality at the GUT scale
within the framework of supersymmetric SU(5) grand unification.
We have taken a model independent approach making use of
mass insertion parameters.
We have examined three patterns of $\ded{23}{LL}$ and $\ded{23}{RR}$:
$LL$, $RR$, and $LL=RR$.
For reconciling $\phi_s$ with LFV,
it greatly helps to choose the optimal value of
the GUT scale gaugino to scalar mass ratio,
in all these three scenarios.
It appears that the most adequate to fit the current value
of $\phi_s$ is $LL=RR$ among the three scenarios. 
The rest two might still be able to push $\phi_s$
into its 90\% CL range.
The barriers to this purpose in the $LL$ scenario are $\bsg$ and $\SphiK$,
but there are cases with low $\tb$ where
they leave a corner of the parameter space satisfying $\phi_s$.
In the $RR$ scenario, the major obstacles are LFV and the neutron EDM\@.
Yet, the former is not totally mutually exclusive with $\phi_s$,
and the latter can be circumvented by a modification
to the $LL$ insertion.
The neutron EDM is a potential danger in the $LL=RR$ scenario as well
depending on the relative phase of the two insertions.
Inclusion of Planck-suppressed non-renormalizable terms for fixing
the quark--lepton mass relations, in general, affects a LFV bound.
This alteration can be estimated by weakening a $\tmg$ bound
to that from $\tau \rightarrow (e + \mu)\,\gamma$.
In the two scenarios involving an $RR$ mixing,
this reduces the tension between LFV and $\phi_s$ for higher $m_0$,
while $\meg$ keeps disfavoring lower $m_0$.
In all cases, low $\tb$ loosens $\bsg$, $\SphiK$, and $d_n$ as well as LFV,
providing for more room to accommodate $\phi_s$.


We thank Diego Tonelli for useful comments.
JhP acknowledges Research Grants funded jointly by the Italian
Ministero dell'Istruzione, dell'Universit\`{a} e della Ricerca (MIUR),
by the University of Padova and
by the Istituto Nazionale di Fisica Nucleare (INFN) within the
\textit{Astroparticle Physics Project} and the FA51 INFN Research Project.
This research was supported in part by the European Community Research
Training Network UniverseNet under contract MRTN-CT-2006-035863.
The work of MY was partially supported by the grants-in-aid from the
Ministry of Education,
Science, Sports and Culture in Japan, No.\ 16081202 and No.\ 17340062.



\end{document}